# Authentication without Identification

# Using

# Anonymous Credential System

Dr. A. Damodaram,

Prof of CSE Dept & Director,

UGC- ASC, JNTUH, Hyderabad,

damodarama@jntuh.ac.in

H.Jayasri,

Asst.Prof,  ATRI, Hyderabad,India,

jayahsree@rediffmail.com

**Abstract**

*Privacy and security are often intertwined. For example, identity theft is rampant because we have become accustomed to authentication by identification. To obtain some service, we provide enough information about our identity for an unscrupulous person to steal it (for example, we give our credit card number to Amazon.com). One of the consequences is that many people avoid e-commerce entirely due to privacy and security concerns. The solution is to perform authentication without identification. In fact, all on-line actions should be as anonymous as possible, for this is the only way to guarantee security for the overall system. A credential system is a system in which users can obtain credentials from organizations and demonstrate possession of these credentials. Such a system is anonymous when transactions carried out by the same user cannot be linked. An anonymous credential system is of significant practical relevance because it is the best means of providing privacy for users.*

## 1.Introduction

As information becomes increasingly accessible, protecting the privacy of individuals becomes a more challenging task. To solve this problem, an application that allows the individual to control the dissemination of personal information is needed. In this paper we discuss about the best known  idea for  such a system called the Anonymous Credential  System.

An anonymous credential(AC) is a vector of attributes certified by a trusted certification authority. Anonymous credentials are used as a way to prevent disclosure of too much information about a user during the authentication process. Access management systems create profiles for each user who has been granted access. Additionally, some systems use digital certificates to further verify the user's identity. Depending on the system, these digital certificate's may contain a lot of information about an individual user's identity. Since the  entire digital certificate  is used during authentication, if compromised it could lead to a breach of sensitive information about the user, some of which could be used later for stealing the legitimate user's identity or authentication credentials for malicious access. The technology was also called minimal disclosure certificates by Stefan Brands.

Here's a scenario to explain how it works. someone goes into a bar and the bartender asks for the person's driver's license to verify if he or she is of legal age to drink. Most likely, the bartender just looks at the person's date of birth and isn't interested in the name, address or other personal information. Once the bartender is satisfied, the person puts their license away and is allowed to stay in the bar. But in open network's -- like the Web and the Internet -- an entire digital certificate may be exposed to the whole world over the wire, where its contents can be sniffed and stolen by hacker's interested in stealing authentication credentials. Minimal disclosure solve that problem by only providing enough information from the user's digital certificate to grant access to a system for a specific requirement. the user's whole identity or credentials aren't served up to the system requesting authentication.

An anonymous credential system consists of users and organizations. Organizations know the users only by pseudonyms. Different pseudonyms of the same user cannot be linked. Yet, an organization can issue a credential to a pseudonym, and the corresponding user can prove possession of this credential to another organization(who knows her by a different pseudonym),without revealing anything more than the fact that she owns such a credential. Credentials can be for unlimited use (multiple-show credentials)or for one-time use(one-show credentials).

## 2.Motivation

The internet, by design, lacks provisions for identifying who communicates with whom; it lacks a well-designed identity infrastructure. As a result, enterprises, governments and individuals have over time developed a bricolage of isolated, incompatible, partial solutions to meet their needs in communications and transactions. The overall result of these unguided developments is that enterprises and governments have a problem in identifying their communication partners at the individual level. Given the lack of a proper identity infrastructure, individuals often have to disclose more personal data than strictly required. In addition to name and address contact details such as multiple phone numbers (home, work, mobile) and e-mail addresses are requested. The amount and nature of the data disclosed exceeds that usually required of real world transactions, which can often be conducted anonymously – in many cases the service could be provided without any personal data at all. Over the long run, the inadequacy of the identity infrastructure affects individuals' privacy. The availability of abundant personal data to enterprises and governments has a profound impact on the individual's right to be let alone as well as on society at large.

## 3. Desirable Properties

### 3.1 Basic Desirable Properties

i) It should be possible for a user to selectively disclose attributes.
ii) An AC must be hard to forge.
iii) A user's transactions must be unlinkable and
iv) An AC must be revokable

### 3.2 Additional Desirable Properties

i) Users should be discouraged from sharing their pseudonyms and credentials with other users. ( PKI assured non- transferability or all-or-nothing non-transferability)
ii) It may be desirable to have a mechanism for discovering the identity of a user whose transactions are illegal.
iii)It can also be beneficial to allow one-show credentials ie, credentials that should only be usable once and should incorporate an offline double spending test.

## 4.Requirements

A basic credential system has users, organizations, and verifiers as types of players. Users are entities that receive credentials. The set of users in the system may grow over time. Organizations are entities that grant and verify the credentials of the users. Each organization grants a unique (for simplicity of exposition) type of credential. Finally, verifiers are entities that verify credentials of the users.

For the purposes of non-transferability, we can add a CA(Certification Authority) to the model who verifies that the users entering the system possess an external public and secret key. This CA will be trusted to do his job properly.

To allow revocable anonymity, an anonymity revocation manager can be added. This entity will be trusted not to use his ability to find out a user's identity or pseudonym unless dictated to do so. The user is anonymous until the revocation manager exposes his/her identity. Usually this is followed by entering the user ID into a revocation list. Revocation may be partial or total. In the former a subset of the entries in the vector is revoked, while in the latter the whole vector is revoked.(ie. the user is revoked.) Ideally revocation authority should not be able to revoke capriciously.

Finally, a credential may include an attribute, such as an expiration date.

## 5. Related Work

The scenario with multiple users who, while remaining anonymous to the organizations, manage to transfer credentials from one organization to another, was first introduced by Chaum [6]. Subsequently, Chaum and Evertse[5] proposed a solution that is based on the existence of a semi-trusted third party who is involved in all transactions. However, the involvement of a semi-trusted third party is undesirable.

The scheme later proposed by Damgard [4] employs general complexity theoretic primitives (one-way functions and zero-knowledge proofs) and is therefore not applicable for practical use. Moreover, it does not protect organizations against colluding users. The scheme proposed by Chen [3] is based on discrete logarithm-based blind signatures. It is efficient but does not address the problem of colluding users. Another drawback of her scheme and the other practical schemes previously proposed is that to use a credential several times, a user needs to obtain several signatures from the issuing organization.

Lysyanskaya, Rivest, Sahai, and Wolf [1] propose a general credential system. While their general solution captures many of the desirable properties, it is not usable in practice because their constructions are based on one-way functions and general zero-knowledge proofs. Their practical construction, based on a non-standard discrete-logarithm-based assumption, has the same problem as the one due to Chen [4]: a user needs to obtain several signatures from the issuing organization in order to use unlinkably a credential several times.

Brands provides a certificate system in which a user has control over what is known about the attributes of a pseudonym. Although a credential system with one-show credentials can be inferred from his framework, obtaining a credential system with multi-show credentials is not immediate and may in fact be impossible in practice. Another inconvenience of these and the other discrete-logarithm-based schemes mentioned above is that all the users and the certification authorities in these schemes need to share the same discrete logarithm group.

Jan Camenisch & Anna Lysyanskaya [2] propose a practical anonymous credential system that is based on the strong RSA assumption and the Diffie-Hellman assumption. They gave the first practical solution that allows 1) a user to unlinkably demonstrate possession of a credential as many times as necessary without involving the issuing organization 2) to prevent misuse of anonymity they offered optional anonymity revocation for particular transaction.

## 6. Concluding Remarks

It appears that a compromise is required, either in the security requirements or in the amount of trust bestowed on the participant, in order to achieve a practical and efficient anonymous credential system.

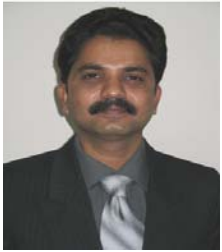

**Dr A Damodaram** obtained his B.Tech. Degree in Computer Science and Engg.in 1989, M.Tech. in CSE in 1995 and Ph.D in Computer Science in 2000 all from Jawaharlal Nehru Technological University, Hyderabad. His areas of interest are Computer Networks, Software Engineering, Data Mining and Image Processing. He presented more than 44 papers in various National and International Conferences and has 7 publications in Journals. He guided 3 Ph.D., 3 MS and more than 100 M.Tech./MCA students.

He joined as Faculty of Computer Science and Engineering in 1989 at JNTU, Hyderabad. He worked in the JNTU in various capacities since 1989. Presently he is a professor in Computer Science and Engineering Department. In his 19 years of service Dr. A. Damodaram assumed office as Head of the Department, Vice-Principal and presently is the Director of UGC Academic Staff College of JNT University Hyderabad.

He was board of studies chairman for JNTU Computer Science and Engineering Branch (JNTUCEH) for a period of 2 years. He is a life member in various professional bodies. He is a member in various academic councils in various Universities. He is also a UGC Nominated member in various expert/advisory committees of Universities in India. He was a member of NBA (AICTE) sectoral committee and also a member in various committees in State and Central Governments. He is an active participant in various social/welfare activities. He was also acted as Secretary General and Chairman for the AP State Federation of University Teachers Associations, and Vice President for All India Federation of University Teachers Associations. He is the Vice President for the All India Peace and Solidarity Organization from Andhra Pradesh.

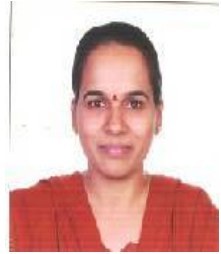

**H.Jayasri** obtained B.E. (CSE) from Bangalore University and M.Tech.(CSE) from JNTU, Hyderabad in 2001 and 2006 respectively. Pursuing Ph.D. from department of CSE JNTU, Hyderabad. She has 7yrs of teaching experience in various colleges of Hyderabad and Bangalore. Areas of research interest are Network Security and Computer Networks.